\documentclass[runningheads]{llncs}
\usepackage[utf8]{inputenc}
\usepackage{graphicx}
\usepackage[hyphens]{url}
\usepackage{hyperref}

\usepackage{todonotes}
\usepackage{cleveref}
\usepackage{microtype}
\usepackage{float}
\usepackage{adjustbox}
\usepackage{rotating}
\usepackage{tikz}

\begin{document}
\title{Backstabber's Knife Collection: A Review of Open Source Software Supply Chain Attacks}
\titlerunning{Backstabber's Knife Collection}

%
\author{Marc Ohm\inst{1} \and
Henrik Plate\inst{2} \and
Arnold Sykosch\inst{1,3} \and
Michael Meier\inst{1,3}
}
\authorrunning{M. Ohm et al.}
%
\institute{University of Bonn, Institute for Computer Science 4,\\ Endenicher Allee 19A, 53115 Bonn, Germany \\
\email{\{ohm,sykosch,mm\}@cs.uni-bonn.de}  \and
SAP Labs France, 805 Av. Maurice Donat, 06250 Mougins, France \\
\email{henrik.plate@sap.com} \and
Fraunhofer FKIE, Department for Cyber Security,\\ Zanderstraße 5, 53177 Bonn, Germany}

\maketitle

\begin{abstract}
A software supply chain attack is characterized by the injection of malicious code into a software package in order to compromise dependent systems further down the chain.
Recent years saw a number of supply chain attacks that leverage the increasing use of open source during software development, which is facilitated by dependency managers that automatically resolve, download and install hundreds of open source packages throughout the software life cycle.

This paper presents a dataset of 174 malicious software packages that were used in real-world attacks on open source software supply chains, and which were distributed via the popular package repositories npm, PyPI, and RubyGems.
Those packages, dating from November 2015 to November 2019, were manually collected and analyzed.
The paper also presents two general attack trees to provide a structured overview about techniques to inject malicious code into the dependency tree of downstream users, and to execute such code at different times and under different conditions.

This work is meant to facilitate the future development of preventive and detective safeguards by open source and research communities.
\end{abstract}


\section{Introduction}%
\label{sec:introduction}
In general, software supply chain attacks aim to inject malicious code into a software product.
Frequently, attackers tamper with the end product of a given vendor such that it carries a valid digital signature, as it is signed by the respective vendor, and may be obtained by end-users through trusted distribution channels, e.g.\ download or update sites.

A prominent example of such supply chain attacks is NotPetya, a ransomware concealed in a malicious update of a popular Ukrainian accounting software~\cite{NotPetya}.
In 2017, NotPetya targeted Ukrainian companies but also hit global corporations, caused damage worth billions of dollars and is said to be one of the most devastating cyberattacks known today~\cite{NotPetyaRussia}.
In the same year, a malicious version of CCleaner, a popular maintenance tool for Microsoft Windows systems, was downloadable from the vendor's official website, and remained undetected for more than a month.
During this period it was downloaded around 2.3 million times~\cite{CCleaner}.

Another flavor of supply chain attacks aims at injecting the malicious code into a dependency of a software vendor's product.
This attack vector was already predicted by Elias Levy in 2003~\cite{levy2003poisoning}, and recent years saw a number of real-world attacks following that scheme.
Such attacks become possible, because modern software projects commonly depend on multiple open source packages, which themselves introduce numerous transitive dependencies~\cite{baker2019keep}.
Such attacks abuse the developers' trust in the authenticity and integrity of packages hosted on commonly used servers and their adoption of automated build systems that encourage this practice~\cite{Fortify2007}.

A single open source package may be required by several thousands of open source software projects~\cite{janaszek2018state}, which makes open source packages a very attractive target for software supply chain attacks.
A recent attack on the npm package \texttt{event-stream} demonstrates the potential reach of such attacks: The alleged attacker was granted ownership of a prominent npm package simply by asking the original developer to take over its maintenance.
At that time, \texttt{event-stream} was used by another 1,600 packages, and was in average downloaded 1.5 million times a week~\cite{event-stream-2018}.

Open source software supply chain attacks are comparable to the problem of vulnerable open source packages which may pass their vulnerability to dependent software projects.
This is known as one of the OWASP Top-10 application security risks~\cite{OWASP2017Top10}.
However, in case of supply chain attacks, malicious code is deliberately injected and attackers employ obfuscation and evasion techniques to avoid detection by humans or program analysis tools.

The major contributions of this paper are as follows: First, the manual analysis and categorization of a dataset with malicious code from 174 packages that were used for real-world attacks on open source software supply chains between 2015 and 2019.
Second, two attack trees that abstract the structure of such attacks, and which were developed both on the basis of the dataset and by reviewing and investigating potential attacks on and actual weaknesses of open source ecosystems.
Through the highlighting of possible entry points, the contributions uncover protection requirements, and serve the further research and development of detective and protective measures against supply chain attacks.

The remainder of the paper is structured as follows:
\Cref{sec:relatedwork} summarizes related work and \Cref{sec:methodology} outlines the methodology used for the two major contributions of this paper.
\Cref{sec:attackmodel} presents the two attack trees and \Cref{sec:dataset} presents the analysis and categorization of the actual code of 174 malicious packages observed in the wild.
\Cref{sec:conclusions} summarizes and concludes the paper.

\section{Related Work}%
\label{sec:relatedwork}
Related work mostly covers \emph{vulnerable} packages, which contain design flaws or code errors that are accidentally introduced, without bad intention but through negligence, and which may pose a potential security risk.
In contrast to that, \emph{malicious} packages contain design flaws or code errors that have been implemented selectively, with caution and the intention to be exploited or triggered at later times in the software life cycle.
Technically, malicious code and vulnerable code may look identical, the main difference lies in the intention of the developer (or lack thereof) and, in some cases, the use of evasion or obfuscation techniques to hinder the detection of such code.

Malicious and vulnerable packages reside in the same ecosystem and live through the same software life cycle.
As such, related works that investigate package reuse in open source ecosystems in general, or the impact and spread of vulnerable packages in particular, also apply to malicious packages.

Decan, Mens, and Constantinou leveraged security reports in order to examine how and when vulnerabilities in npm software packages are discovered and fixed.
In order to assess the effect on other packages hosted on npm, a dependency graph was used.
The key findings are that nearly half of the packages inherited vulnerabilities from other packages, and that version pinning to vulnerable and outdated packages are the main cause for such inherited vulnerabilities, even if fixes are available.~\cite{decan2018impact}

Zimmermann, Staicu, Tenny, and Pradel were able to verify these findings and provide mitigation techniques.
Highly popular packages and highly active developers were identified as single point of failures.
Thus, the authors propose to raise developer awareness through training as well as automated code analysis.~\cite{zimmermann2019small}

Pfretzschner and Othmane proposed a system to identify software supply chain attacks in npm packages by static code analysis.
The tool is able to detect four kinds of attacks: Leakage of global variables, manipulation of global variables, local function manipulation, and dependency-tree manipulation.
However, the authors failed to identify real-world examples of these attacks for evaluation.~\cite{pfretzschner2017identification}

Garrett, Ferreira, Jia, Sunshine, and K{\"a}stner proposed anomaly detection through unsupervised learning in order to identify suspicious package updates.
For that purpose they collected over 700,000 packages from npm and normal behavior was inferred from 1,500 randomly selected packages.
The system reported 539 suspicious updates per week reducing manually inspection by 89\%.~\cite{garrett2019detecting}

Jukka Ruohonen examined vulnerable Python packages regarding their CVSS (Common Vulnerability Scoring System) score and the respective weakness according to CWE (Common Weakness Enumeration).
An auto regressive model was used to calculate how likely a new release is vulnerable based on previous releases' vulnerability.
It was found that the prediction of this event is difficult despite good statistical performance.
However, the supply chain of a package was not taken into consideration.~\cite{ruohonen2018empirical}

While related work mostly focused on \emph{vulnerable} packages and impact assessment with regard to \emph{specific open source ecosystems}, especially Node.js (npm), this work considers \emph{malicious} packages \emph{across several ecosystems}.

\section{Methodology}%
\label{sec:methodology}
It is important to distinguish between vulnerable and malicious packages.
As said, \emph{vulnerable} packages may contain design flaws or code errors that are accidentally introduced, without bad intention but through negligence, and which may pose a potential security risk.
According to the Cambridge Dictionary \emph{malicious} means ,,\textbf{intended} to cause damage to a computer system, or to steal private information from a computer system''.
Technically, malicious and vulnerable coding can be similar or even identical, thus, the main difference lies in the attacker's intention.
The contribution of this work is two-fold: A systematic description of possible attacks using attack trees, and a dataset of malicious packages used in real-world attacks.

Attack trees are able to represent attacks against a system.
The primary goal of the attack is used as root node and child nodes represent possible ways to achieve that goal.
The attack trees described in Section~\ref{sec:attackmodel} have been created in an iterative fashion on the basis of actual malicious packages observed in the wild, and on the basis of potential attacks and weaknesses described by security researchers and practitioners.
The first attack tree has the objective to inject malicious code into software supply chains of downstream users, while the second aims at triggering its malicious behavior in different circumstances.
Each malicious package and information source is manually analyzed and, provided sufficient information is available, mapped to a node of each attack tree.
If no fitting node is present, a new one is added.
This iterative approach makes sure that the nodes of the attack trees represent realistic attack vectors.
However, the approach has the disadvantage that those trees are not complete, as they do not reflect attack vectors that have not been observed or otherwise described.

The second contribution is the dataset.
It is analyzed in Section~\ref{sec:dataset} and comprises the subset of malicious packages used in real-world attacks for which the actual malicious code could be obtained (typically a compressed archive).
The compilation took place between July 2\textsuperscript{nd} and August 2\textsuperscript{nd}, 2019 and was updated on 27\textsuperscript{th} of January 2020.
The programming languages JavaScript with its package repository npm, Java (Maven Central), Python (PyPI), PHP (Packagist) and Ruby (RubyGems), which are the most popular languages according to newly created GitHub repositories in 2018~\cite{elliott2018toplanguages}, are covered by the dataset.

During that time, the vulnerability database Snyk\footnote{\url{https://snyk.io}}, language-specific security advisories, and research blogs were reviewed to identify malicious packages and possible attack vectors.
It must be noted that these sources solely mention the packages' names and affected versions, thus,
the actual malicious code has to be downloaded from other sources.
However, such malicious packages are typically not available anymore on standard package repositories of the respective programming language, e.g.\ npm or PyPI.
Instead, where possible, they were retrieved from deprecated mirrors, internet archives, and public research repositories.
If the code of a malicious package could be retrieved, it was analyzed and categorized manually.
This was done in order to confirm the packages' maliciousness, map them to the existing attack trees or extend them if necessary.
The publication of malicious versions of a package are dated according to Libraries.io\footnote{\url{https://libraries.io}}, a service that monitors package releases across all major package repositories.
Advisories and public incident reports are used to date the public disclosure of the malicious package.


\section{Threat Analysis and Attack Trees}%
\label{sec:attackmodel}
This section starts with a high-level introduction of activities and systems related to open source software development projects, and concludes with the presentation of two attack trees.

In general, attack trees allow for a systematic description of attacks against any kind of system\footnote{\url{https://www.schneier.com/academic/archives/1999/12/attack_trees.html}}. The root node of a given tree thereby corresponds to the attacker's top-level goal, and child nodes represent alternative ways to achieve it. The top-level goals of the attack trees presented in Sections~\ref{sec:tree1} and \ref{sec:tree2} are to inject malicious code into the software supply chain, thus, into a dependency of a development project, and to trigger that malicious code in different circumstances.

\subsection{Open Source Development Projects}
\begin{figure}[tb]
  \centering
  \includegraphics[width=\linewidth]{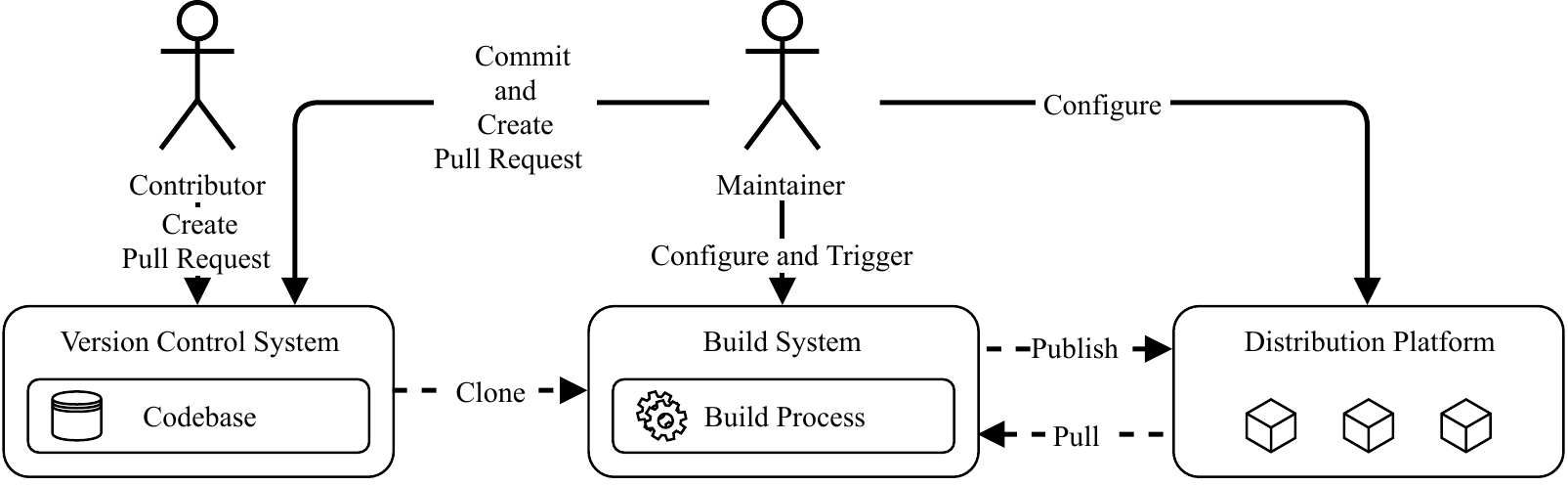}
  \caption{High-level development and build activities.}%
  \label{fig:build-process}
\end{figure}

In a typical development environment as visualized in \Cref{fig:build-process}, \emph{Maintainers} are members of a development project who administer the depicted systems, provide, review and approve contributions, or define and trigger build processes.
Open source projects also receive code contributions from \emph{contributors}, which may be reviewed and merged into the project's code base by maintainers.

The \emph{build process} ingests the source code and other resources of a project, and has the goal to produce software artifacts.
These artifacts are subsequently published such that they become available to end-users and other development projects.

The project resources reside in a \emph{version control system} (VCS), e.g.\ Git, and are copied to the local file system of the \emph{build system}.
Among those resources is a declaration of direct dependencies, which is analyzed at the start of the build process by a \emph{dependency manager} in order to establish the complete dependency tree with all direct and transitive dependencies.
As all of them are required during the build, for instance, at compile time or during test execution, they are downloaded (pulled) from \emph{package repositories} such as PyPI\footnote{\url{https://pypi.org}} for Python, npm\footnote{\url{https://www.npmjs.com}} for Node.js, or Maven Central\footnote{\url{https://search.maven.org/}} for Java.

At the end of a successful build, program code and other resources are assembled into one or more build artifacts, which are potentially signed and eventually published.
Either to distribution platforms like app stores such that they may be consumed by end-users or to \emph{package repositories} for other development projects.

Such project environments are subject to numerous trust boundaries, and many threats target the respective data flows, data stores and processes.
Managing those threats may be challenging even when considering only the environment of a single software project.
When considering supply chains with dozens or hundreds of dependencies, it is important to notice that such an environment exist for every single dependency, making it obvious that the combined attack surface of such projects is considerably larger than that of software entirely developed in-house.

Taking the perspective of attackers, malevolent actors have the intention to compromise the security of the build or runtime environment of software projects through the infection of one or more upstream open source packages, each one of which is developed in environments comparable to \Cref{fig:build-process}.
How to reach this goal is described in the following sections by means of two attack trees that provide a structured overview about attack paths to inject a malicious code into dependency trees of downstream users and to trigger its execution at different times or under different conditions.

\subsection{Injection of Malicious Code}
\label{sec:tree1}
\begin{figure}[tb]
  \centering
  \includegraphics[width=\textwidth]{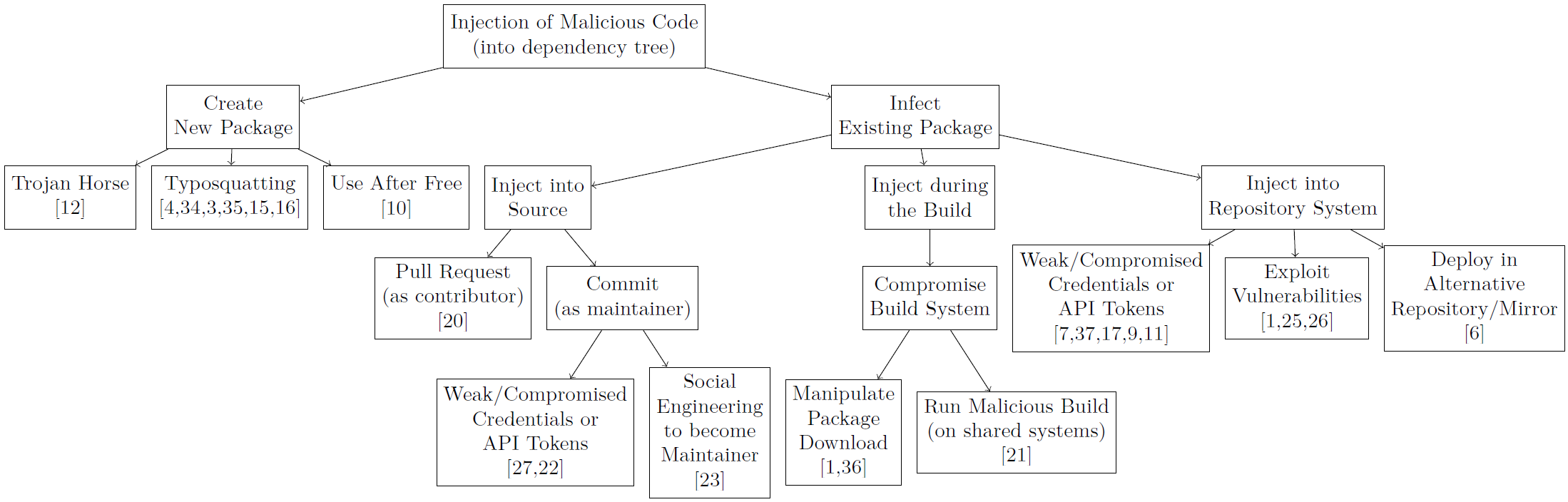}
  \caption{Attack tree to inject malicious code into dependency trees.}%
  \label{fig:tree-1}
\end{figure}

The attack tree illustrated by \Cref{fig:tree-1} is an extension and refinement of the graph presented by Pfretzschner and Othmane \cite{pfretzschner2017identification}, and has as top-level goal to inject malicious code into the dependency tree of downstream packages.
Thus, the goal is satisfied once a package with malicious code is available on a distribution platform, e.g.\ package repository, and it became a direct or transitive dependency of one or more other packages.
As such, this type of code injection differs from other injection attacks, many of which exploit security vulnerabilities at application runtime, e.g.\ buffer overflow attacks that become possible due to a lack of user input sanitization.
%

To inject a package into dependency trees an attacker may follow two possible strategies, he may either \emph{infect an existing package} or submit a \emph{new package}.

Obviously, developing and publishing a new rogue package using a name that is not used by anybody else avoids interference with other legitimate project maintainers.
However, such a package has to be discovered and referenced by downstream users in order to end up in the dependency trees of victim packages.
This may be achieved using a name similar to existing package names (\emph{typosquatting})~\cite{pypi-squatting-2019,40npm-typo,pypi-colourama-crypto-miner,Tschacher2016,denbraver2019jellyfish,dunn2017acquisition}, or by developing and promoting a \emph{trojan horse}~\cite{npm-getcookies-backdoor-2018}.
An attacker might also use the opportunity to reuse the identifier of an existing project, package, or user account withdrawn by its original and legitimate maintainer (\emph{use after free})~\cite{gh-account-reuse-2018}.

The second strategy is to infect an \emph{existing package} that already has users, contributors and maintainers.
The attacker might choose packages for different reasons, e.g.\ a significant number or specific group of downstream users.
However, the data gathered until now does not allow to validate corresponding hypotheses (c.f.\ \Cref{sec:dataset}).
Once the attacker choose a package to infect, the malicious code may be injected \emph{into the sources}, \emph{during the build}, or \emph{into the package repository}.

Open source projects live and strive through community contributions, thus, attackers can mimic benign project contributors.
For instance, an attacker may pretend to solve an existing issue by \emph{creating a pull request} (PR) with a bug fix or a seemingly useful feature or dependency~\cite{CCskimming2018}.
The latter could be used to create a dependency on an attacker-controller package created from scratch using the techniques described beforehand.
In any case, this PR has to be approved and merged into the main code branch by a legitimate project maintainer.
Alternatively, an attacker may \emph{commit} malicious code into the project's code base all by himself by using \emph{weak or compromised credentials} or security-sensitive API tokens~\cite{gentoo2018,homebrew2018}.
Furthermore, attackers may become maintainer themselves through \emph{social engineering}~\cite{event-stream-2018}.
In all cases, no matter how the malicious code has been added to the sources, it will become part of an official package during the next release build --- regardless where that build happens.
Compared to attacks on build systems and package repositories, malicious code in VCS is more accessible to manual or automated reviews of commits or entire repositories.

The \emph{compromise of build systems} typically entails tampering with resources used throughout the build process, e.g.\ compilers, build plugins or network services such as proxies or DNS servers.
Such resources may be compromised if the build system, be it a developer's work station or a build server like Jenkins\footnote{\url{https://jenkins.io}}, is subject to vulnerabilities, or if insecure communication channels are such that attackers can \emph{manipulate the package download} from repositories~\cite{Fortify2007,mvn-http-proxy-2013}.
The release builds of the targeted package may also run on a shared build system and thus used by multiple projects~\cite{Gruhn:2013:SPC:2491055.2491070}.
Depending on the setup, such build processes may not run in isolation, hence resources such as package caches or build plugins are shared between the builds of different projects. In this case, an attacker may compromise shared resources during a \emph{malicious build} of a project under his control such that the targeted project is compromised at a later point in time.

Even popular package repositories are still subject to simple but severe security vulnerabilities.
While all the other attack vectors seek to inject malicious code into a single package, the \emph{exploit of vulnerabilities} in package repositories themselves puts the entire repository with all its packages at risk~\cite{RubyGems-2017,packagist-2018}.
Similar to injecting the code in the sources, the attacker may use \emph{weak or compromised credentials}~\cite{weak-npm-creds-2017,eslint-2018,edge2019backdoor-bootstrap-sass,ssh-decorator-2018,coe2018conventional} or gain maintainer authorizations through \emph{social engineering}~\cite{event-stream-2018} in order to publish malicious versions of legitimate packages.
As the former has been used in numerous attacks, initiatives such as the badge program of the Core Infrastructure Initiative \cite{CII2019vulnerability} give the official recommendation to project maintainers to enable two-factor authentication.

Further, an attacker may upload malicious package versions to \emph{alternative repositories or repository mirrors}~\cite{android-audio-recorder-2018,android-audio-recorder-2018-2} that are not provisioned by the original maintainers, and wait for victims pulling dependencies from there.
Supposedly, such repositories and mirrors are less popular, and the attack is dependent on the victim's configuration, e.g.\ the order of repositories queried for dependencies or the use of mirrors.

\subsection{Execution of Malicious Code}
\label{sec:tree2}
\begin{figure}[tb]
  \centering
  \includegraphics[width=\textwidth]{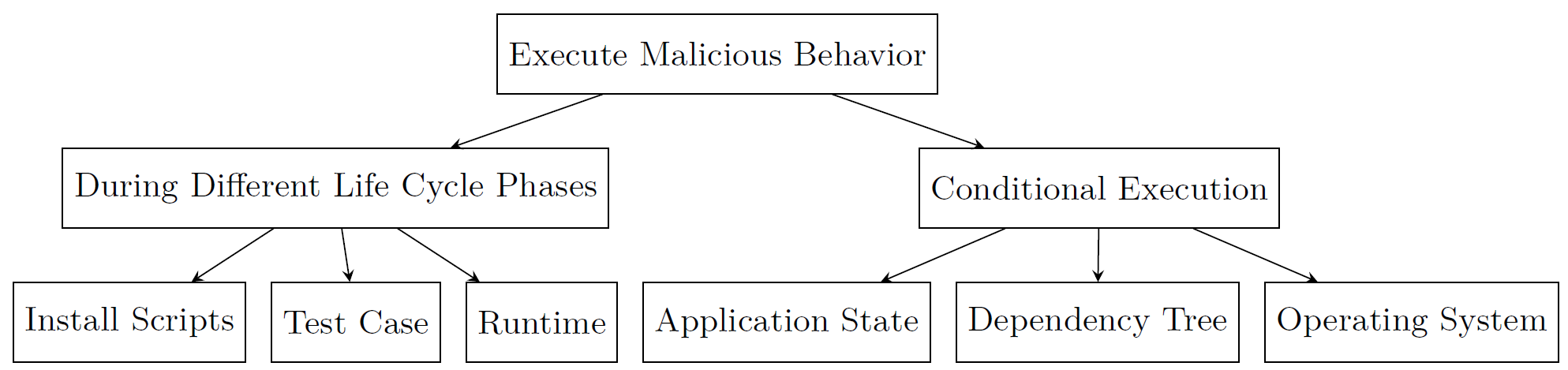}
  \caption{Attack tree to execute malicious code.}%
  \label{fig:tree-2}
\end{figure}

Once malicious code is present in some project's dependency tree, the attack tree illustrated by \Cref{fig:tree-2} has the top-level goal to trigger the malicious code under different conditions.
Such conditions may be used to evade detection and/or target attacks towards specific users and systems.

Malicious code may trigger at different \emph{life cycle phases} of the infected package and its downstream users (c.f.\ \Cref{sec:trigger-of-malicious-behavior}).
If malicious code is contained in \emph{test cases}, the attack primarily targets the contributors and maintainers of the infected package, which run such tests on their developer work stations and build systems.
In many of the recorded attacks, malicious code is contained in \emph{install scripts}, which are automatically executed during package installation by downstream users or their dependency managers.
Such install scripts exist for Python and Node.js, and may be used to perform pre- or post-installation activities.
Malicious code in install scripts puts the contributors and maintainers of downstream packages as well as their end users at risk.
Malicious code may also be triggered at \emph{runtime} of downstream packages, which requires that it is invoked as part of the regular control flow of the victim package.
In Python, this may be achieved by including malicious code in \texttt{\_\_init\_\_.py}, which is invoked through \textit{import} statements.
In JavaScript, this may be achieved by monkey-patching (modifying) existing methods.
The specifics of individual programming languages, package managers, etc.\ may easily be covered by refining this goal.

Independent of the life cycle phase, the execution of malicious behavior may always trigger (unconditioned) or only if certain conditions are met (\emph{conditional execution}).
As for any other malware, conditioned execution complicates the dynamic detection of malicious open source packages, since the respective conditions may not be known, understood or met in sandbox environments.
Conditioning the execution on the \emph{application state} is a common means to evade detection, e.g.\ in test environments or dedicated malware analysis sandboxes.
Again, the specifics of individual build systems may be covered by respective sub goals, e.g.\ the presence of Jenkins environment variables indicates that malicious code is triggered during a build rather than in a production environment.
Moreover, conditions may be related to a specific victim package, e.g.\ check a specific application state such as the balance of a crypto wallet~\cite{event-stream-2018}.
Heavy reuse of open source packages may result in the fact that a malicious package ends up in the dependency tree of many downstream packages.
If only certain packages are of interest to attackers, they may condition the code execution on the nodes of a given \emph{dependency tree} at hand~\cite{event-stream-2018}.
Furthermore, the \emph{operating system} used may serve as condition.

\section{Description of the Dataset}%
\label{sec:dataset}
The dataset contains 174 packages and was compiled according to our methodology as described in \Cref{sec:methodology}.

A total number of 469 malicious packages could be identified.
Additionally, 59 packages were found that could be identified as proof of concept (published by researchers) and hence are excluded from further examination.
Eventually, we were able to obtain at least one affected version for 174 packages.
The rate of successful downloads of malicious packages for npm is 109/374 (29.14\%), for PyPI 28/44 (63.64\%), for RubyGems 37/41 (90.24\%), and for Maven Central 0/10 (0.00\%).
All statements and statistics below refers to the set of downloaded packages.

\subsection{Composition and Structure}%
\label{sec:composition-and-structure}
The dataset consist of 62.6\% packages published on npm and hence are written for Node.js in JavaScript.
The remaining packages were published via PyPI (16.1\%, Python) and via RubyGems (21.3\%, Ruby).
Unfortunately, a malicious Java package targeting Android developers could not be downloaded.
For PHP, we were not able to identify any malicious package at all.

The complete dataset is available for free on GitHub\footnote{https://dasfreak.github.io/Backstabbers-Knife-Collection}.
However, access will be granted on justified request only due to ethical reasons.
The dataset is structured as follows: \textit{package-manager/package-name/version/package.file}.
Malicious packages are grouped by their originating package manager on the first level.
Further, multiple affected versions of one package are grouped under the respective package's name.
As example for the affected version of the well-known case of event-stream it is: \textit{npm/event-stream/3.3.6/event-stream-3.3.6.tgz}.

\begin{figure}[t]
  \centering
  \begin{minipage}{.48\textwidth}
    \includegraphics[width=\textwidth]{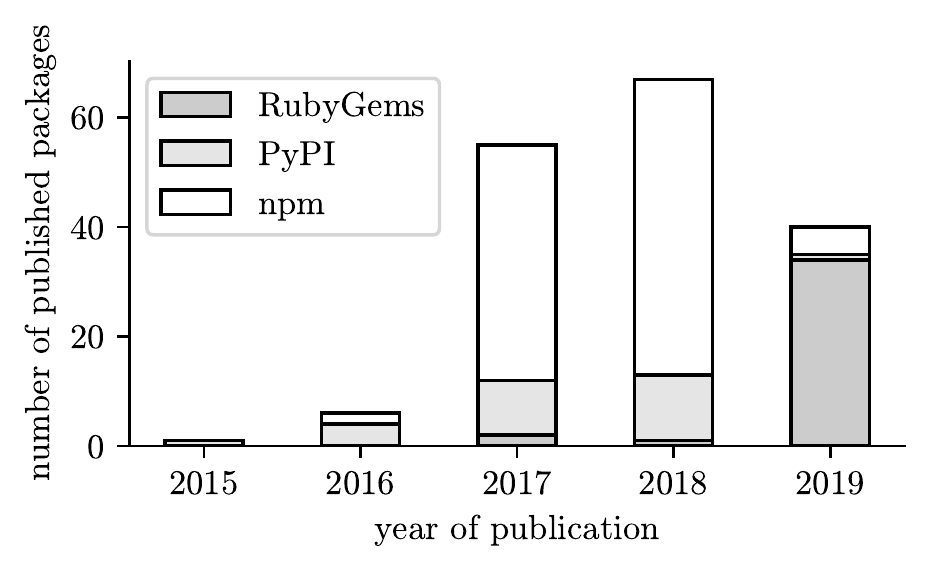}
    \caption{Publication dates of collected packages.}%
    \label{fig:published}
  \end{minipage}
  \begin{minipage}{.48\textwidth}
    \includegraphics[width=\textwidth]{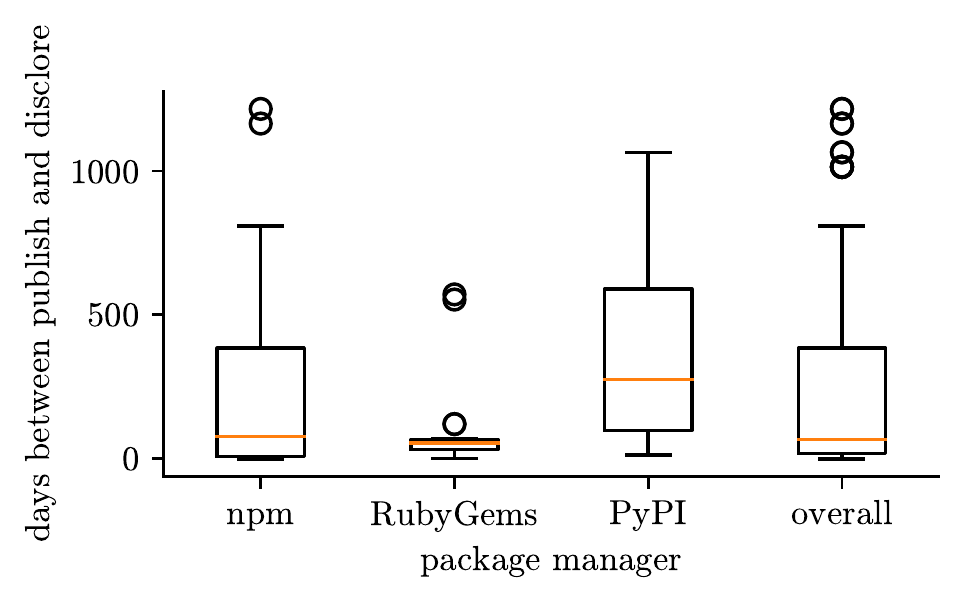}
    \caption{Temporal distance between date of publication and disclosure.}%
    \label{fig:diff-publish-reported}
  \end{minipage}
\end{figure}

\subsection{Temporal Aspects}%
\label{sec:temporal-aspects}
\Cref{fig:published} visualizes the publication dates of the collected packages which range from November 2015 to November 2019.
The publication and disclosure dates are identified according to the upload time of the package and the publication date of the corresponding advisory identifying the respective version as malicious (cf. \Cref{sec:methodology}).
A trend for an increasing number of published malicious packages is apparent.
While malicious packages for PyPI are known to date back to 2015 and since then are increasing, npm gained a massive amount of malicious packages in 2017, and malicious packages on RubyGems experienced a boom in 2019.

\Cref{fig:diff-publish-reported} shows that \textbf{on average a malicious package is available for 209 days} $(min=-1, max=1,216, \rho=258, \tilde{x}=67)$ before being publicly reported.
A minimum of $-1$ days was reached for \textit{npm/eslint-config-airbnb-standard/2.1.1} which was affected by \textit{npm/eslint-scope/3.7.2}.
Even though the infection of \textit{npm/eslint-scope/3.7.2} was known, the package was still in use due to the developers' re-packaging strategy.
The maximum of $1,216$ days was reached by \textit{npm/rpc-websocket/0.7.7} which took over an abandoned package and went undetected for a long period.

In general this shows that packages tend to be available for a longer period.
While PyPI has the highest average online time, that period varies the most for npm, and RubyGems tends to detect malicious packages more timely.

\subsection{Trigger of Malicious Behavior}%
\begin{figure}[t]
\centering
  \includegraphics[width=\textwidth]{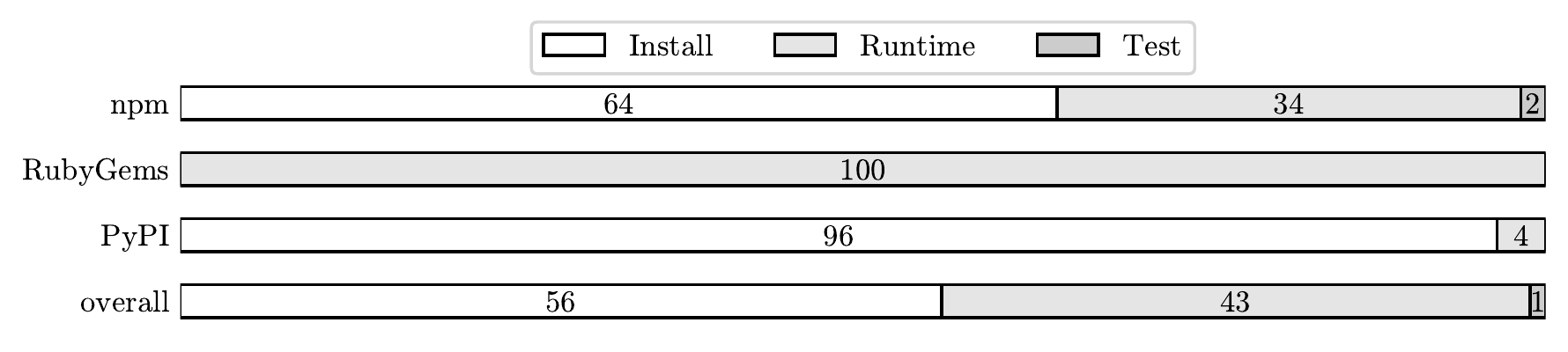}
  \caption{Trigger of malicious behavior separated per package repository and overall.}%
  \label{fig:trigger}
\end{figure}

\label{sec:trigger-of-malicious-behavior}
Malicious behavior of a package may be triggered at different points of interaction with the package.
Most typically, a package may be installed, tested, or executed.
A separation per package repository is visualized in \Cref{fig:trigger}.
It illustrates that the poor handling of arbitrary code during install yields the most used infection vector.

It is apparent that \textbf{most malicious packages (56\%) start their routines on installation}.
This can be triggered by the package repositories' install command, e.g.\ \texttt{npm install <package>}.
That invokes code as defined in the package's definition, e.g.\ \texttt{package.json} and \texttt{setup.py}.
This code might be arbitrary to do whatever is necessary to install the package, e.g.\ download additional files.
This seems very common for malicious packages on PyPI.

In contrast to that, Ruby does not implement such install logic and hence no packages for that case exist in Ruby.
Thus, all found packages on RubyGems use runtime as trigger.
Overall, 43\% of the packages expose their malicious behavior during the program's runtime, i.e.\.\ when invoked from another function.

For 1\% of the packages the test routines are used as trigger.
Invoking the test routine of \textit{npm/ladder-text-js/1.0.0} would execute \texttt{sudo rm -rf /*} which may delete all your files.

\subsection{Conditional Execution}%
\begin{figure}[tb]
  \centering
  \includegraphics[width=\textwidth]{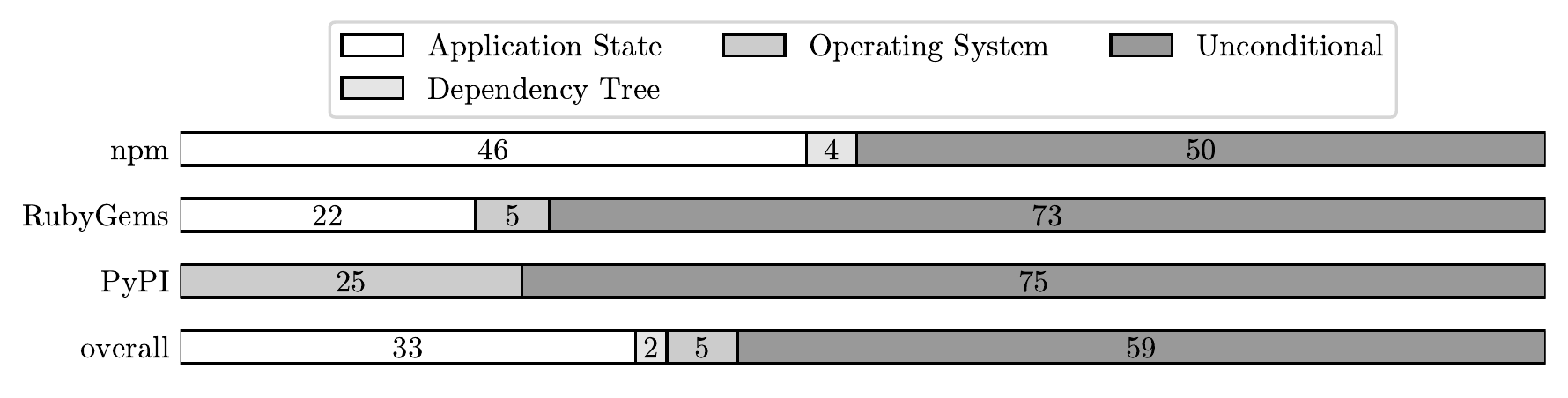}
  \caption{Ratio of conditional and unconditional execution per package repository and overall.}%
  \label{fig:conditional}
\end{figure}

\label{sec:conditional-execution}
As seen in \Cref{fig:conditional}, \textbf{41\% of the packages check for a condition before triggering further execution}.
This may depend on the application's state, e.g.\ check whether the main application is in production mode (e.g.\ \textit{RubyGems/paranoid2/1.1.6}), resolvability of a domain name (e.g.\ \textit{npm/logsymbles/2.2.0}), or the amount contained in a crypto wallet (e.g.\ \textit{npm/flatmap-stream/0.1.1}).

Other techniques are to check whether another package is present in the dependency tree (e.g.\ \textit{npm/load-from-cwd-or-npm/3.0.2}) or whether the package is executed on a certain operating system (e.g.\ \textit{PyPI/libpeshka/0.6}).

The majority of packages published on PyPI and RubyGems execute unconditionally.
For npm the ratio of conditional and unconditional execution is nearly equal.

\subsection{Injection of Malicious Package}%
\begin{figure}[b]
  \centering
  \includegraphics[width=\textwidth]{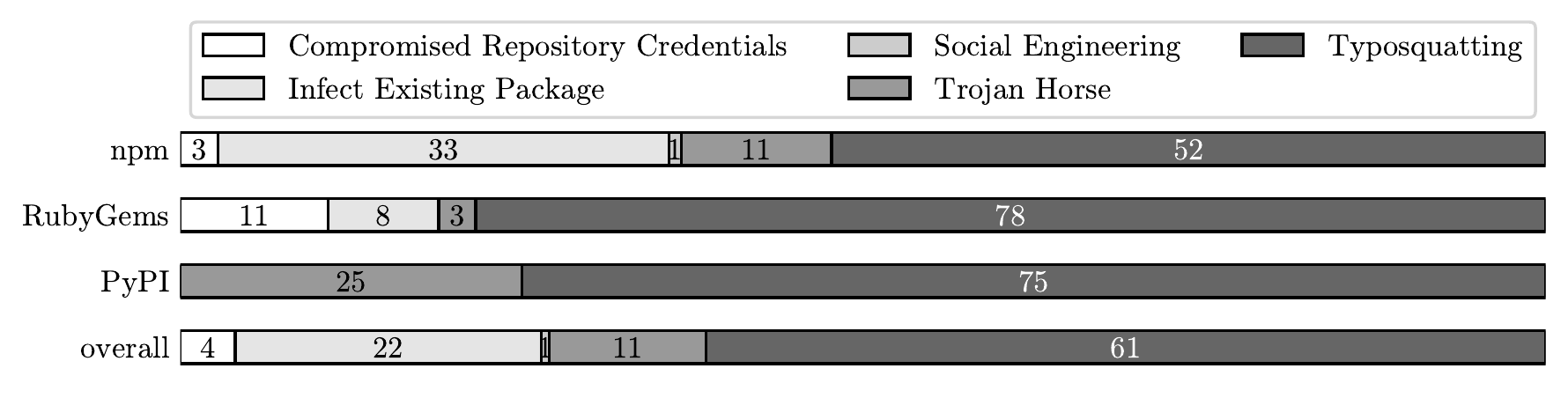}
  \caption{Injection technique used to introduce the malicious package into a package per package repository and overall.}%
  \label{fig:injection}
\end{figure}

\label{sec:injection-of-malicious-package}
In \Cref{fig:injection} it is apparent that \textbf{most (61\%) malicious packages mimic existing packages' names via typosquatting}.
A deeper analysis of that phenomenon revealed that the Levenshtein distance of an average typosquatting package to its target is 2.3 $(min=0, max=11, \rho=2.05, \tilde{x}=1.0)$.
In some cases the typosquatting target is available from another package repository, e.g.\ the Linux package repository \texttt{apt} under the exact same name.
This is for instance the case for \textit{python-sqlite}.
The maximum distance of $11$ is reached in the case of \textit{pythonkafka} which targeted \textit{kafka-python}.
Common techniques are adding or removing hyphens, leaving out single letters, or exchange of letters that are often mistyped.
An word that is targeted exceptionally often is ,,color'' or, respectively, its British English counterpart: ,,colour''.

The second most common injection method was the infection of an existing package.
This may often be achieved with \emph{compromised credentials} for the repository system (e.g.\ \textit{npm/eslint-scope/3.7.2}).
In most cases, the exact infection technique could not be determined in retrospect.
This is because the related source is often removed from the version control system or no further details about the injection are made public.
Hence, these packages are listed as \emph{infect existing package}.

Another injection technique is to create a new package which consist of nothing but the malicious package to which we refer to as \emph{trojan horse}.
No meaningful typo-squatting targets where found for these packages.
These packages could be used in conjunction with an infected existing package or standalone.

\subsection{Primary Objective}%
\begin{figure}[H]
  \centering
  \includegraphics[width=\textwidth]{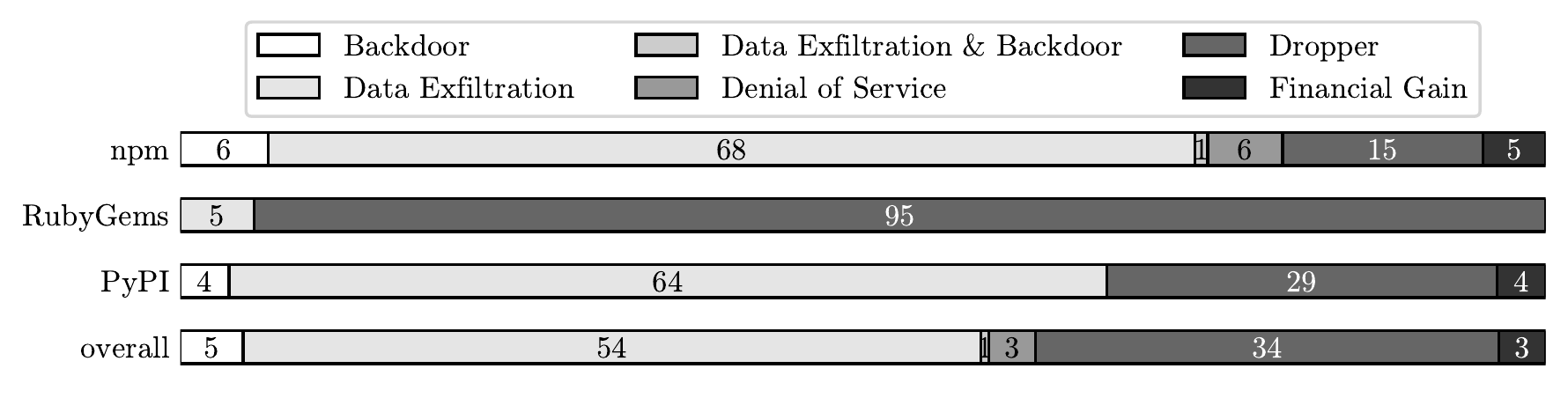}
  \caption{Primary objective of the malicious package per package repository and overall.}%
  \label{fig:objective}
\end{figure}

\label{sec:primary-objective}
As shown in \Cref{fig:objective}, \textbf{most packages aim at data exfiltration}.
Commonly, the data of interest is the content of \texttt{/etc/passwd}, \texttt{$\mathtt{\sim}$/.ssh/*}, \texttt{$\mathtt{\sim}$/.npmrc}, or \texttt{$\mathtt{\sim}$/.bash\_history}.
Furthermore, malicious packages try to exfiltrate environment variables (which might contain access tokens) and general system information.
Another popular target (7 reported packages, 3 of them available in our dataset) is the token for the voice and text chat application Discord.
A Discord user's account may be linked to credit card information and thus be used for financial fraudulence.

Moreover, 34\% of the packages function as Dropper to download second stage payload.
Another 5\% open a backdoor, i.e.\.\ reverse shell, to a remote server and await further instructions.
3\% aim to cause a denial of service by exhausting resources through fork bombs and file deletion (e.g.\ \textit{npm/destroyer-of-worlds/1.0.0}) or breaking functionality of other packages (e.g.\ \textit{npm/load-from-cwd-or-npm/3.0.2}).
Only 3\% have financial gain as primary objective by for instance running a cryptominer in the background (e.g.\ \textit{npm/hooka-tools/1.0.0}) or stealing cryptocurrency directly (e.g.\ \textit{pip/colourama/0.1.6}).
In addition, combinations of the above mentioned objectives might occur.

\subsection{Targeted Operating System}%
\label{sec:targeted-operating-system}
In order to identify the targeted OS, the source code was manually analyzed for hints which may be as explicit as an if–then construct like \texttt{if platform.system() is 'Windows'} as used in e.g.\ \textit{PyPI/openvc/1.0.0} or implicit by relying on resources only available on certain OS.
These resources may be for instance files containing sensible information like \texttt{.bashrc} etc. (cf. \Cref{sec:primary-objective}, \textit{npm/font-scrubber/1.2.2}) or executables like \texttt{/bin/sh} (e.g.\ \textit{npm/rpc-websocket/0.7.11}).

The analysis of the packages for their targeted operating system (OS) revealed that \textbf{most packages (53\%) are agnostic, i.e.\.\ do not rely on OS-specific functions}.
The analysis was done on the initial visible code of the package and thus the targeted OS of the second stage payload remains unknown.
However, Unix-like systems seem to be targeted more often than Windows and macOS, since build environments are commonly operated on such an OS.

There is only one known case of macOS being the target in which the package \textit{npm/angluar-cli/0.0.1} performs a denial of service attack on the McAfee virus scanner for macOS by deleting and modifying its files.

\begin{figure}[t]
  \centering
  \includegraphics[width=\textwidth]{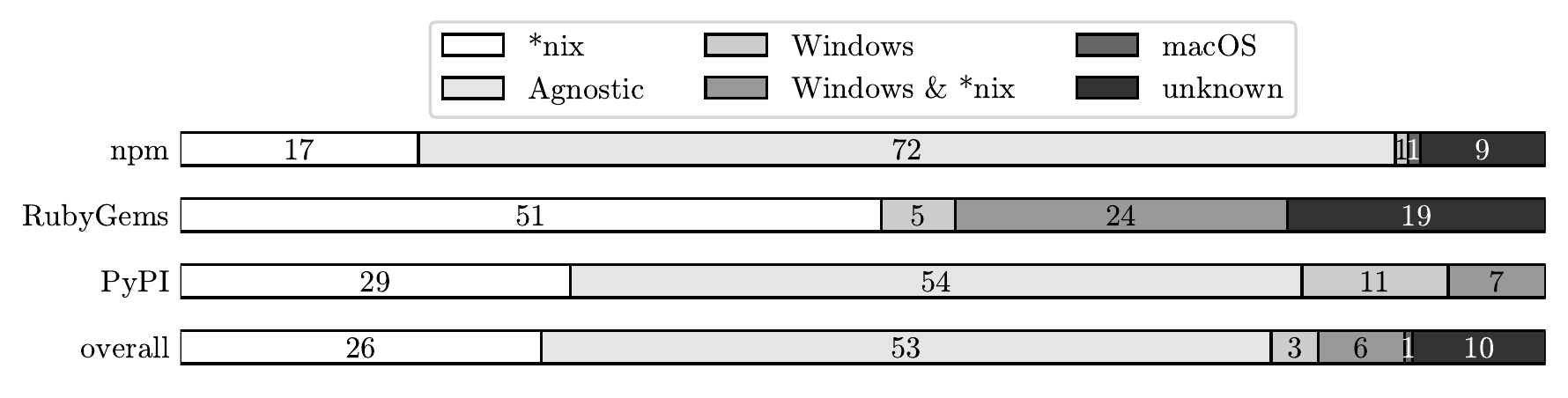}
  \caption{Targeted operating system per package repository and overall.}%
  \label{fig:target-os}
\end{figure}

\subsection{Obfuscation}%
\label{sec:obfuscation}
Malicious actors often try to disguise the presence of their code, i.e.\.\ hindering its detection by sight.
In our dataset \textbf{nearly the half of the packages (49\%) employ some kind of obfuscation}.
Most often a different encoding (Base64 or Hex) is used to disguise the presence of malicious functions or suspicious variables such as domain names.

A technique often used by benign packages to compress source code and thus save bandwidth is minification.
However, this is a welcome opportunity for malicious actors to sneak in extra code which is unreadable for humans (e.g.\ \textit{npm/tensorplow/1.0.0}).
Another way to hide variables is to use string sampling.
This requires a seemingly random string which is used to rebuild meaningful strings by picking letter by letter (e.g.\ \textit{npm/ember-power-timepicker/1.0.8}).

In one case the malicious functions are hidden by encryption.
The package \textit{npm/flatmap-stream/0.1.1} leverages AES256 with the short description of the targeted package as decryption key.
That way, the malicious behavior is solely exposed when used by the targeted package.
Furthermore, combinations of the above mentioned techniques exist.

\begin{figure}[tb]
  \centering
  \includegraphics[width=\textwidth]{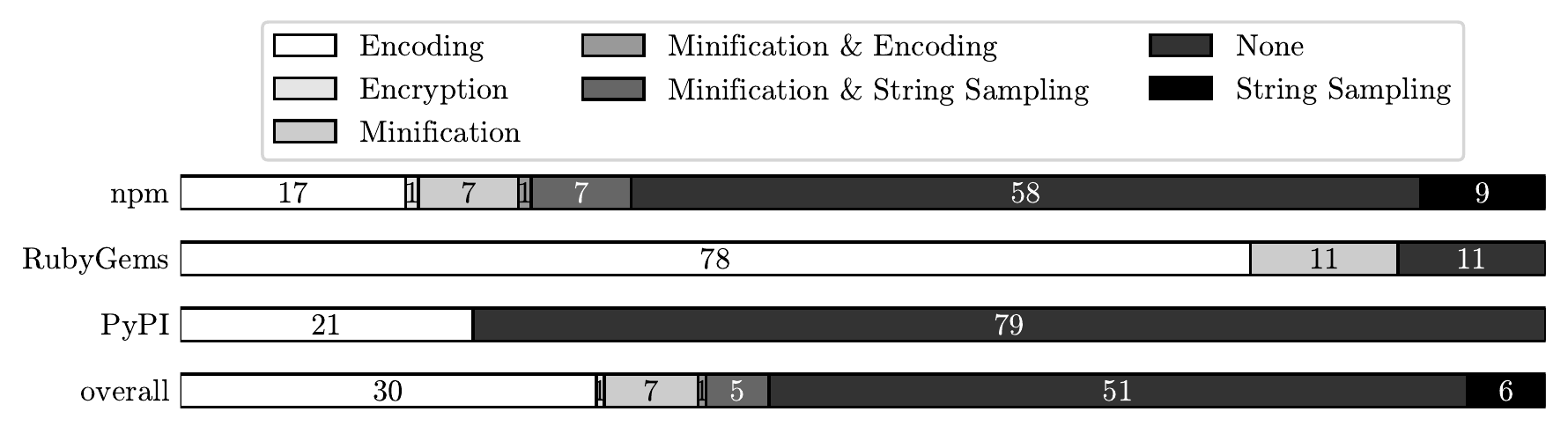}
  \caption{Employed obfuscation technique per package repository and overall.}%
  \label{fig:obfuscation}
\end{figure}

\subsection{Clusters}%
\label{sec:clusters}
In order to infer on the presence of attack campaigns, all packages were analyzed for reuse of malicious code or dependency relationships.
This way, \textbf{it was possible to identify 21 clusters} for which at least two packages either have similar malicious code in common, or an attacker-controlled package depends on another one with the actual malicious code. In total, 157 of the 174 packages (90\%) belong to a cluster.
On average a cluster comprises 7.28 packages $(min=2, max=36, \rho=8.96, \tilde{x}=3)$.

A cross comparison of publications dates of packages within one cluster revealed that the average temporal distance between publications is 42 days, 6:50:18 $(min=\textrm{1:29:40}, max=\textrm{353 days, 11:17:02}, \rho=\textrm{78 days, 0:43:10}, \tilde{x}=\textrm{7 days, 15:24:51})$.
The biggest cluster was formed around the \texttt{crossenv} case~\cite{40npm-typo} counting 36 packages published with an average temporal distance of 5.98 days.
It was published in two waves, 11 packages within 15 minutes on 19\textsuperscript{th} of July 2017 and another 25 packages within 30 minutes on 1\textsuperscript{st} of August 2017.

The cluster having publication dates that are 353 days apart consists of the two packages \textit{PyPI/jeilyfish/0.7.0} and \textit{PyPI/python3-dateutil/2.9.1}.
The first was published on 12/11/18 12:26 AM and contained code that download a script to steal SSH and GPG Keys from Windows machines.
It went undetected for a long time until the second package was published on 11/29/19 11:43 AM which did not contain malicious code itself but referenced the first package.
The cluster was reported and deleted on 12/12/19 05:53 PM.

While most clusters solely contain packages from one package repository, it was possible to find a cluster that mainly contained packages from npm but also \textit{RubyGems/active-support/5.2.0} from RubyGems. This means that attack campaigns exist or at least techniques flow across multiple package repositories.

\subsection{Code Review of Two Malicious Packages}
\begin{figure}[p]
  \begin{adjustbox}{addcode={\begin{minipage}{\width}}{\caption{%
      Comparison and explanation of two malicious packages: \textit{npm/jqeury/3.3.1} on the left; \textit{RubyGems/active-support/5.2.0} on the right. The code was found in \texttt{ext/trellislike/unflaming/waffling/extconf.rb} and \texttt{package/src/data/var/everted.js}, respectively. Names of classes, variables, and function are randomly picked from an English dictionary (different for each package of the cluster). After instantiation (1), a DNS lookup (2) for an obfuscated (Base64) domain is started. Both snippet contain code for conditional execution. If the lookup is either empty or the non-routable meta-address \texttt{0.0.0.0} (3), the execution of malicious code is skipped. This may be seen as evasion technique as it may be the case for sandboxed environments as used for dynamic malware analysis. However, if the domain can be resolved, a HTTP GET request is send (4). The response is written to new a file located in \texttt{/tmp} (5, 6) and executed (7).}
      \label{fig:notable-example}\end{minipage}},rotate=90,center}
      \includegraphics[width=\textheight]{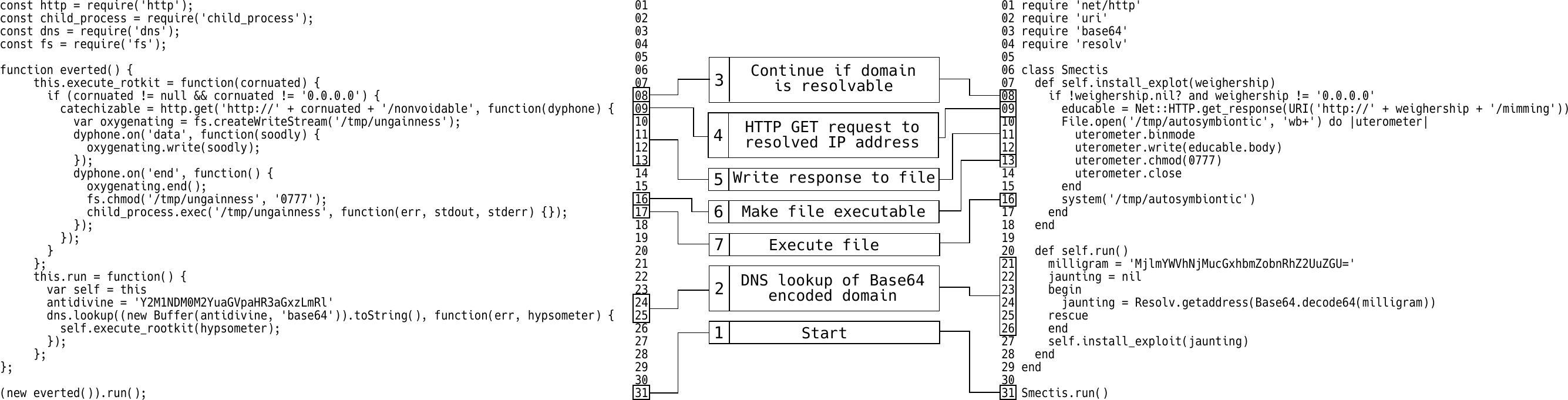}
  \end{adjustbox}
\end{figure}

For vivid illustration, \textit{npm/jqeury/3.3.1} (left) and \textit{RubyGems/active-support/5.2.0} (right) will be discussed in \Cref{fig:notable-example}.
They both belong to the same cluster according to our manual assessment of code similarity, even though they were published on different repositories.

\section{Conclusions}%
\label{sec:conclusions}
From an attacker’s point of view, package repositories represent a reliable and scalable malware distribution channel.
So far, the repositories of Node.js (npm) and Python (PyPI) are the primary targets of malicious packages, supposedly due to the fact that malicious code can be easily triggered during package installation.
There already exist a number of countermeasures that may be implemented by different stakeholders, e.g.\ multi-factor authentication for open source maintainers, version pinning and disablement of install scripts for open source users, or the isolation of build processes and hardening of build servers.
However, despite raising general awareness among stakeholders, such countermeasures must be more accessible and, where possible, enforced by default in order to prevent open source software supply chain attacks.
The following paragraphs briefly summarize and reflect our findings and future work.

\subsection{Findings}%
\label{sec:findings}
Two attack trees were derived from observed cases and related work.
One for the injection of a malicious package into the open source ecosystem and one for the execution of the malicious code.
These attack trees allow for systematic description of past and future attacks.
We were able to create the first manually curated dataset of malicious open source packages that have been used in real-world attacks.
It consists of 174 malicious packages (62.6\% npm, 16.1\% PyPI, 21.3\% RubyGems) ranging from November 2015 to November 2019.
Manual analysis revealed that most packages (56\%) trigger their malicious behavior on installation, and 41\% use further conditions to determine whether to run or not.
More than half of the packages (61\%) leverage typosquatting to inject themselves into the ecosystem, and data exfiltration is the most common goal (55\%).
The packages typically are agnostic to operating systems (53\%), and often employ obfuscations (49\%) to hide themselves.
Finally, we were able to detect multiple clusters of malicious packages through reused code even across different programming languages.
The dataset gives insight and is available for free to facilitate research in the area of prevention, detection, and mitigation of software supply chain attacks.

\subsection{Limitation}%
\label{sec:limitations}
Our dataset is highly biased towards malicious packages that are written in JavaScript for Node.js and published on npm which is due to npm's enormous size and popularity.
Unfortunately, we were not able to obtain malicious packages for Java (Maven Central) and PHP (Packagist).
Furthermore, roughly 34\% of the malicious packages are droppers with the goal to download a second stage payload, which might not be available anymore.
One might notice that we listed the deployment in alternative repository or mirrors as injection method but downloaded most of the packages from such sources.
While it is possible that these packages have been altered to be malicious, the package's presence in our dataset is still valid as the package would be malicious is both cases.
Furthermore, the ,,intended'' maliciousness according to the advisories was verified through manual analysis.

\subsection{Future Work}%
\label{sec:futurework}
It will be important to collect a comprehensive set of existing safeguards, and perform a gap analysis with regard to the attack vectors described in \Cref{sec:attackmodel}.
Even though many safeguards already exist, some are hardly used in practice, and others will need to be developed.
For example, we expect new techniques and tools to scan entire package repositories for suspicious packages, e.g.\ on the basis of the observation that malicious code is reused across packages of the same campaign, and even across languages.
In this context, the manually curated and labeled dataset allows for supervised learning approaches that support the automated and repository-wide search for malicious packages.
Moreover, with regard to existing and new mitigation strategies, the presented dataset may pose as a benchmark.
Last, acknowledging the importance of a comprehensive and up-to-date dataset, it will be necessary to continue its curation -- contributions are welcome.

\bibliographystyle{splncs04}
\bibliography{references}

\end{document}